\documentclass[aps,prd,superscriptaddress,twocolumn]{revtex4}

\usepackage{amsfonts}

\usepackage{amsmath}
\usepackage{graphicx}
\usepackage{color}

\DeclareMathAlphabet\mathbfcal{OMS}{cmsy}{b}{n}

\begin{document}

\title{Testing Lorentz- and CPT-invariance with ultracold neutrons}

\author{A. Mart\'{i}n-Ruiz}
\email{alberto.martin@nucleares.unam.mx}
\affiliation{Instituto de Ciencia de Materiales de Madrid, CSIC, Cantoblanco, 28049 Madrid, Spain}
\affiliation{Centro de Ciencias de la Complejidad, Universidad Nacional Aut\'{o}noma de M\'{e}xico, 04510 Ciudad de M\'{e}xico, M\'{e}xico}

\author{C. A. Escobar}
\email{cruiz@ualg.pt}
\affiliation{CENTRA, Departamento de F\'{i}sica, Universidade do Algarve, 8005-139 Faro, Portugal}

\begin{abstract}
In this paper we investigate, within the standard model extension framework, the influence of Lorentz- and CPT-violating terms on gravitational quantum states of ultracold neutrons. Using a semiclassical wave packet, we derive the effective nonrelativistic Hamiltonian which describes the neutrons vertical motion by averaging the contributions from the perpendicular coordinates to the free falling axis. We compute the physical implications of the Lorentz- and CPT-violating terms on the spectra. The comparison of our results with those obtained in the GRANIT experiment leads to an upper bound for the symmetries-violation $c_{\mu\nu} ^{n}$ coefficients. We find that ultracold neutrons are sensitive to the $a _{i} ^{n}$ and $e _{i} ^{n}$ coefficients, which thus far are unbounded by experiments in the neutron sector. We propose two additional problems involving ultracold neutrons which could be relevant for improving our current bounds; namely, gravity-resonance-spectroscopy and neutron whispering gallery wave.
\end{abstract}

\maketitle

\section{Introduction}

One of the main challenges of modern physics is the search for a quantum theory of gravity (QTG). On the experimental front, the major difficulty is the lack of experimentally accessible phenomena at Planck scale that could shed light on a possible route to QTG. However, suppressed effects emerging from the underlying theory might be observable in sensitive experiments performed at our presently low-energy scales. One candidate set of Planck scale signals is relativity violations, which are associated with the breaking of Lorentz and CPT symmetries, hence the considerable amount of attention it has gained in the past two decades. Some modern approaches to QTG, such as noncommutative field theories \cite{noconmutativo}, quantum gravity \cite{qg1}, string theory \cite{string1}, brane-worlds scenarios \cite{brane1}, condensed matter analogues of ``emergent gravity'' \cite{cmatter}, Ho\v{r}ava-Lifshitz gravity \cite{horava1,Petrov1,Petrov2}, gauge emergent bosons \cite{boson1} and others Lorentz-violating scenarios \cite{others,others2,others3}, are examples that lead to setups in which Lorentz invariance is no longer an exact symmetry.

Studies of Lorentz violation (LV) are conducted more easily in low-energy effective field theory frameworks, which allow us to focus on measurable physical effects rather than the fundamental mechanism that produces the breakdown of Lorentz symmetry. In particular, the Standard-Model Extension (SME) \cite{SME1a} was conceived within these low-energy frameworks to encompass all possible LV effects. The Lagrangian of the minimal SME include the standard model and general relativity terms plus all the Lorentz-violating operators of mass dimension four or less that can be constructed from the coupling of the standard fields with vector and tensor coefficients that parameterize Lorentz violation. Such coefficients are motivated by a spontaneous symmetry breaking in a more fundamental theory \cite{String1ab} and whose fixed directions in spacetime trigger the breakdown of Lorentz symmetry. It is worth to mention that some properties as observer Lorentz invariance, energy-momentum conservation, gauge invariance, power-counting renormalizability \cite{14a,14b}, causality, stability and hermiticity (see \cite{17a} for the fermion sector and \cite{Potting2} for the photon sector) can be maintained in the Lagrangians of the minimal SME.

Since Lorentz violation has not been detected yet in experiments, it is generally assumed that LV coefficients have small components in Earth-based laboratories, thus leading to very tiny modifications in physically measurable quantities. There are also some cases where the SME terms lead to new effects which are absent in the Lorentz-symmetric theory, for instance, forbidden decays \cite{Ralfd}, magnetoelectric phenomena \cite{magne1} and birefringence in vacuum \cite{birre}. High precision experiments have been used to find tighter bounds for the LV coefficients (see Ref. \cite{table1} for current bounds). For example, the SME causes small shifts in the energy levels of an atomic system that could, in principle, be detected by high-precision spectroscopy. This idea has been used to set stringent bounds to the electron sector of the SME, since the 2S-1S transition in hydrogen has been measured with particularly high precision \cite{Hydrogen}. The neutron sector of the SME has received less attention and current bounds on the Lorentz- and CPT-violating coefficients are based mainly on nuclear binding models and Cs interferometers.

In this paper we consider the physics of ultracold neutrons (UCNs) as a possible candidate to test Lorentz- and CPT-invariance. In particular, the recently observed gravitational quantum states of UCNs in the GRANIT experiments \cite{GRANIT} offer an interesting opportunity of testing departures from both the neutrons quantum mechanical behavior and possible modifications of the local gravity field \cite{Bertolami}. This fact motivates the investigation of SME effects on neutron gravitational quantum states, which is precisely the question we address here. A detailed description of the GRANIT experiment at the Institute Lau-Langevin can be found in Ref. \cite{GRANIT}. In short, they show that an intense beam of UCNs moving in Earth's gravity field does not bounce smoothly but at certain well-defined quantized heights, as predicted by quantum theory. Since we aim to compare our theoretical results with the ones obtained in the GRANIT experiments, we frame this work according to the laboratory conditions under which experiments were carried out. To this end, we start with the fermion sector of the SME coupled to a general curved spacetime background and we work out its spin-independent nonrelativistic expansion, which is appropriate to describe the dynamics of an unpolarized beam of slow neutrons. Since the neutrons motion in the plane perpendicular to the free falling axis is governed by classical laws, we use a Gaussian wave packet to derive an effective Hamiltonian which describes the SME effects on the quantum bouncer. The resulting energy shifts can be compared with the results obtained in the GRANIT experiments, and an upper bound can be set for the Lorentz- and CPT-violating coefficients.

This paper is organized as follows. We begin in Section \ref{Generalities} by introducing the nonrelativistic Hamiltonian which describes the spin-independent effects of a nonrelativistic fermion in a uniform Newtonian gravitational field. We closely follow Ref. \cite{Yuri}, wherefrom we take notations and conventions. In Sec. \ref{SecEffHam} we derive the effective Hamiltonian which affects the neutrons motion along the free falling axis. We have relegated the technical computations to the Appendix. Comparing the energy shifts induced by Lorentz violation and the experimental precision in the GRANIT experiments, we set bounds to the $c _{\mu \nu} ^{n}$ SME coefficients in Sec. \ref{BoundSec}. Finally, in Sec. \ref{Summary} we briefly discuss two experiments involving UCNs which can be used to improve our bound to the SME coefficients.

\section{Lorentz violation in a uniform gravitational field} \label{Generalities}

In order to investigate the SME effects on nonrelativistic quantum systems in a uniform Newtonian gravitational field, we have to consider first the action for a single fermion $\psi$ of mass $m$ in a general curved spacetime background. The appropriate SME action is given by \cite{Kostelecky}
\begin{align}
S = \int e \left[ \frac{i}{2} e ^{\mu} _{\phantom{\mu} a} \left( \overline{\psi} \Gamma ^{a} \nabla _{\mu} \psi - \left( \nabla _{\mu} \overline{\psi} \right) \Gamma ^{a} \psi \right) - \overline{\psi} M \psi \right] d ^{4}x , \label{Action}
\end{align}
where $e$ is the determinant of the vierbein $e ^{\mu} _{\phantom{\mu}a}$, and the covariant derivative $\nabla _{\mu}$ acts on the spinors as
\begin{align}
\nabla _{\mu} \psi &= \partial _{\mu} \psi + \frac{i}{4} \omega _{\mu} ^{\phantom{\mu} ab} \, \sigma _{ab} \, \psi ,  \\ \nabla _{\mu} \overline{\psi} &= \partial _{\mu} \overline{\psi} - \frac{i}{4} \omega _{\mu} ^{\phantom{\mu} ab} \,  \overline{\psi} \, \sigma _{ab} ,
\end{align}
being $\omega _{\mu} ^{\phantom{\mu} ab}$ the spin connection and $\sigma ^{ab} = \frac{i}{2} \left[ \gamma ^{a} , \gamma ^{b} \right]$. The Dirac matrices $\gamma ^{a}$ are taken to satisfy $\left\lbrace \gamma ^{a} , \gamma ^{b} \right\rbrace = - 2 \eta ^{ab}$, where $\eta ^{ab} =$ diag $(-1,1,1,1)$ is the tangent-space metric.

The symbols $\Gamma ^{a}$ and $M$ appearing in the action (\ref{Action}) are defined by
\begin{align}
\Gamma ^{a} \equiv & \; \gamma ^{a} - c _{\mu \nu} e ^{\nu a} e ^{\mu} _{\phantom{\mu} b} \gamma ^{b} - d _{\mu \nu} e ^{\nu a} e ^{\mu} _{\phantom{\mu} b} \gamma _{5} \gamma ^{b} \notag \\ & - e _{\mu} e ^{\mu a} - i f _{\mu} e ^{\mu a} \gamma _{5} - \frac{1}{2} g _{\lambda \mu \nu} e ^{\nu a} e ^{\lambda} _{\phantom{\lambda} b} e ^{\mu} _{\phantom{\mu} c} \sigma ^{bc}   \label{Gamma}
\end{align}
and
\begin{align}
M \equiv m + a _{\mu} e ^{\mu} _{\phantom{\mu} a} \gamma ^{a} + b _{\mu} e ^{\mu} _{\phantom{\mu} a} \gamma _{5} \gamma ^{a} + \frac{1}{2} H _{\mu \nu} e ^{\mu} _{\phantom{\mu} a} e ^{\nu} _{\phantom{\nu} b} \sigma ^{ab} , \label{Mass}
\end{align}
where $\gamma ^{5} = i \gamma ^{0} \gamma ^{1} \gamma ^{2} \gamma ^{3}$. The first term in Eq. (\ref{Gamma}) leads to the usual Lorentz-invariant kinetic term for the Dirac field, while the first term of Eq. (\ref{Mass}) corresponds to the Lorentz-invariant mass. The Lorentz breaking coefficients, $a _{\mu}$, $b _{\mu}$, $c _{\mu \nu}$, $d _{\mu \nu}$, $e _{\mu}$, $f _{\mu}$, $g _{\lambda \mu \nu}$ and $H _{\mu \nu}$, are assumed to have small components in an Earth-based laboratory (concordant frame \cite{17a}).

In any static spacetime the vierbein can be written as $e ^{\mu} _{\phantom{\mu} 0} = \delta ^{\mu} _{0} e ^{0} _{\phantom{0} 0} (x ^{k})$ and $e ^{\mu} _{\phantom{\mu} j} = \delta ^{\mu} _{i} e ^{i} _{\phantom{i} j} (x ^{k})$, where $e ^{0} _{\phantom{0} 0} \neq 0$ \cite{Wald}. The Dirac equation that results from the action (\ref{Action}) can be written as
\begin{align}
i e ^{0} _{\phantom{0} 0} \Gamma ^{0} \partial _{0} \psi =& - i e ^{i} _{\phantom{i} j} \Gamma ^{j} \partial _{i} \psi - \frac{i}{2} e ^{\mu} _{\phantom{\mu}a} \left( \partial _{\mu} \Gamma ^{a} \right) \psi + M \psi \notag \\ &- \frac{i}{2} e ^{\mu} _{\phantom{\mu}a} \omega _{\mu cd} \left( \eta ^{ac} \Gamma ^{d} + \frac{i}{4} \left\lbrace \Gamma ^{a} , \sigma ^{cd} \right\rbrace \right) \psi . \label{DiracEq}
\end{align}
The Hamiltonian $H$ associated with this Dirac equation must satisfy $H \psi = i \partial _{0} \psi$, and this is naively achieved by inverting $e ^{0} _{\phantom{0} 0} \Gamma ^{0}$. However, the resulting Hamiltonian is not hermitian and then is physically unacceptable. As shown in Ref. \cite{Kostelecky}, this problem can be repaired by making a spacetime-constant field redefinition $\psi = W \chi$, where $W$ is chosen to restore the usual time-derivative coupling. In the present case, to first order in the SME coefficients, the hermitian operator $W = \left( 3 - \gamma ^{0} \Gamma ^{0} \right) / 2$ correctly works \cite{Yuri}. The modified Dirac equation takes the standard form $i \partial _{0} \chi = H \chi$, where the hermitian Hamiltonian reads
\begin{align}
H = - i \frac{e ^{i} _{\phantom{i}j}}{e ^{0} _{\phantom{0}0}} \gamma ^{0} \tilde{\Gamma} ^{j} \partial _{i} + \frac{1}{e ^{0} _{\phantom{0}0}} \gamma ^{0} \tilde{M} , \label{Hamiltonian}
\end{align}
with $\bar{W} = \gamma ^{0} W ^{\dagger} \gamma ^{0}$, $\tilde{\Gamma} ^{a} = \bar{W} \Gamma ^{a} W$, and
\begin{align}
\tilde{M} = & \, \bar{W} M W - i e ^{\mu} _{\phantom{\mu}a} \bar{W} \gamma ^{a} \left( \partial _{\mu} W \right) - \frac{i}{2} e ^{\mu} _{\phantom{\mu}a} \bar{W} \left( \partial _{\mu} \Gamma ^{a} \right) W \notag \\ & - \frac{i}{2} e ^{\mu} _{\phantom{\mu}a} \omega _{\mu cd} \bar{W} \left( \eta ^{ac} \Gamma ^{d} + \frac{i}{4} \left\lbrace \Gamma ^{a} , \sigma ^{cd} \right\rbrace \right) W . \label{NewMass}
\end{align}
To proceed further, we have to choose properly the background spacetime in order to characterize the gravitational field in any experiment intended to measure Lorentz violation effects in the fermion sector of the SME. This is achieved by working with the usual uniform Newtonian field, which is described by the vierbein $e ^{\mu} _{\phantom{\mu}\nu} = \delta ^{\mu i} \delta _{\nu j} + \delta ^{\mu 0} \delta _{\nu 0}  \left( 1 + \Phi \right) ^{-1}$, where $\Phi$ is the uniform Newtonian potential \cite{Rasel}. In this case, the resulting relativistic Hamiltonian is \cite{Yuri}
\begin{align}
H = - i \left( 1 + \Phi \right) \gamma ^{0} \tilde{\Gamma} ^{i} \partial _{i} + \left( 1 + \Phi \right) \gamma ^{0} \tilde{M} , \label{Hamiltonian2}
\end{align}
where $\tilde{\Gamma} ^{i} = \gamma ^{i} + \Gamma ^{i} + \left[ \gamma ^{0} \gamma ^{i} , \Gamma ^{0} \right] / 2$ and
\begin{align}
\tilde{M} &= m + M - \frac{m}{2} \left\lbrace \gamma ^{0} , \Gamma ^{0} \right\rbrace - \frac{i}{2} \gamma ^{0} \gamma ^{i} \left( \partial _{i} \Gamma ^{0} \right) - \frac{i}{2} \left( \partial _{i} \Gamma ^{i} \right) \notag \\ & \hspace{2cm} - \frac{i}{2} \frac{\left( \partial _{i} \Phi \right)}{1 + \Phi} \left( \gamma ^{i} + \Gamma ^{i} + \gamma ^{0} \gamma ^{i} \Gamma ^{0} \right) . 
\end{align}
The main goal of this paper is to look for signals or possible effects of Lorentz violation in experiments with ultracold neutrons in the presence of the Earth's gravitational field. UCNs have nonrelativistic velocities and can thus be described by the nonrelativistic limit of the Hamiltonian (\ref{Hamiltonian2}), which can be obtained by using the standard Foldy-Wouthuysen (FW) procedure \cite{FW}. The FW method consists in finding a unitary transformation $S$ in the Hilbert space such that the $4 \times 4$ Hamiltonian $\tilde{H} = e ^{iS} H e ^{-iS}$ is $2 \times 2$ block diagonal, where the leading $2 \times 2$ block then represents the desired nonrelativistic Hamiltonian. Performing the FW transformation for the complete Hamiltonian of Eq. (\ref{Hamiltonian2}) is cumbersome and also unnecessary for our purposes because the GRANIT experiment, which is the one with which we want to compare our results, is performed with an unpolarized beam of slow neutrons \cite{GRANIT}. This requires averaging the spin states thus diminishing the effects of any spin-dependent Lorentz-violating coefficient. This is why, in the remainder of this paper, we focus on general spin-independent SME effects, which are associated with the coefficients $a _{\mu}$, $c _{\mu \nu}$ and $e _{\mu}$.

The detailed derivation of the nonrelativistic Hamiltonian using the FW procedure is presented in Refs. \cite{Kostelecky, Yuri}. The resulting Schr\"{o}dinger operator valid to linear order in $\Phi$ and $\partial _{i} \Phi$ is found to be
\begin{align}
H =& \left( m + \mathfrak{a} _{0} - m \mathfrak{c} _{00} - m \mathfrak{e} _{0} \right) \left( 1 + \Phi \right) \notag \\ & + \frac{\eta ^{ij}}{2m} \left[ \mathfrak{a} _{j} - m \left( \mathfrak{c} _{0j} + \mathfrak{c} _{j0} \right) - m \mathfrak{e} _{j} \right] \left( 2 \hat{p} _{i} + \Phi \hat{p} _{i} + \hat{p} _{i} \Phi \right) \notag \\ & + \frac{1}{2m} \left[ \eta ^{ij} \left( 1 - \mathfrak{c} _{00} \right) - 2 \eta ^{il} \eta ^{jm} \mathfrak{c} _{(lm)} \right] \hat{p} _{(i} \left( 1 + \Phi \right) \hat{p} _{j)}  , \label{Schrodinger}
\end{align}
where $\hat{p _{i}} = - i \partial _{i}$ is the momentum operator which, as usual, acts on all objects on its right. In this expression we have defined the coefficients $\mathfrak{a} _{0} = (1 - \Phi) a _{0}$, $\mathfrak{a} _{j} = a _{j}$, $\mathfrak{e} _{0} = (1 - \Phi) e _{0}$, $\mathfrak{e} _{j} = e _{j}$, $\mathfrak{c} _{00} = (1 - 2 \Phi) c _{00}$, $\mathfrak{c} _{0j} = (1 - \Phi) c _{0j}$ and $\mathfrak{c} _{ij} = c _{ij}$, which acquire additional factors depending on the gravitational potential. The indices inside parentheses denote symmetrization with a factor $1/2$. In the limit where $\Phi = 0$, the Hamiltonian (\ref{Schrodinger}) correctly reduces to the one obtained in Ref. \cite{Kostelecky2}. Moreover, it also reduces to the one reported in Ref. \cite{HehlNi} when all the SME coefficients are set to zero. Notice that the previous analysis holds for any fermion (e.g. electron, neutron, etc.). From now on, we focus on the neutron sector of the SME and then we label the LV-coefficients with an additional superfix $n$ to indicate this fact, i.e., $a _{\mu} ^{n}, c _{\mu \nu} ^{n}$ and  $e _{\mu} ^{n}$.

\section{Effective Hamiltonian}
\label{SecEffHam}

In this section we derive the effective Hamiltonian $H _{\mbox{\scriptsize eff}}$ which describes the SME effects on the quantum free fall of UCNs. We first note that, in any fixed frame, the term $\left( \mathfrak{a} _{0} - m \mathfrak{e} _{0} \right) \left( 1 + \Phi \right)$, which is second order in the gravitational potential, can be absorbed into the rest mass $m$ and therefore is not observable; we shall henceforth ignore both terms. Therefore we are left with the effective Hamiltonian
\begin{align}
H _{\mbox{\scriptsize eff}} =& \frac{\gamma _{ij} ^{-}}{2m} \left[ \hat{p} _{i} \hat{p} _{j} + \frac{1}{c ^{2}} \Phi _{,( i} \hat{p} _{j )} \right] + \frac{\gamma _{ij} ^{+}}{2m c ^{2}} \Phi \hat{p} _{i} \hat{p} _{j} + m \alpha ^{+} \Phi \notag \\ & - \beta _{i} \left( \hat{p} _{i} c + \frac{1}{2c} \Phi _{, i} \right) + \varsigma _{i} \hat{p} _{i} c + \frac{\varsigma _{i}}{c} \left( \Phi \hat{p} _{i} + \frac{1}{2 } \Phi _{,i} \right) \label{EffHam}
\end{align}
where we have used the commutator $\left[ \hat{p} _{i} , \Phi \right] = \Phi _{,i}$, with $\Phi _{, i} \equiv \hat{p} _{i} \Phi$ and $\hat{p} _{i} = - i \hbar \partial _{i}$. In Eq. (\ref{EffHam}) we have restored the fundamental constants $c$ and $\hbar$; and we have defined
\begin{align}
\alpha ^{\pm} \equiv 1 \pm c _{00} ^{n} , \quad \gamma _{ij} ^{\pm} \equiv \delta _{ij} \alpha ^{\pm} -  c _{ij} ^{n} -  c _{ji} ^{n} , \notag \\ \beta _{i} \equiv c _{i0} ^{n} + c _{0i} ^{n} , \quad \varsigma _{i} \equiv (a _{i} ^{n} / m) - e _{i} ^{n} . \label{Constants}
\end{align}
In order to reduce further the Hamiltonian (\ref{EffHam}), let us recall the experimental work performed at the Institute Laue-Langevin by V. V. Nesvizhevsky and coworkers. The GRANIT experiment shows that UCNs moving in the Earth's gravity do not move smoothly but jump from one height to another, as predicted by quantum theory \cite{GRANIT}. In practice, they use an intense horizontal beam of UCNs directed slightly upwards and allowing the neutrons to fall onto a horizontal mirror. By placing a neutron absorber above the mirror and counting the particles as they moved the absorber up and down, they found that neutrons are measured only at certain well-defined heights. In this situation, the horizontal motion of neutrons is governed by classical laws, while the vertical motion is quantized. Ideally, the vertical and horizontal motions of a neutron are independent; however, in a Lorentz-violating background this statement is not longer valid, as we can see in the Hamiltonian (\ref{EffHam}). Based on the above, in this paper we consider that the neutron's motion in the tangent plane to the Earth's surface, which is classical, can be modeled by a Gaussian wave packet of the form
\begin{align}
\psi (\textbf{r} _{\perp}) = \frac{1}{\sqrt{\pi} \sigma} e ^{\frac{i}{\hbar} \textbf{p} _{\perp} \cdot \textbf{r} _{\perp} - \frac{\textbf{r} _{\perp} ^{2}}{2 \sigma ^{2}}} , \label{WavePacket}
\end{align}
where $\textbf{r} _{\perp} = (x,y)$ and $\textbf{p} _{\perp} = (p _{x} , p _{y})$ are the coordinates and momentum in the plane perpendicular to the free fall motion, respectively. The classicality condition requires the characteristic width $\sigma$ of the wave packet to be very small. Since the GRANIT experiment measures the neutrons vertical position, in the following we use the ansatz (\ref{WavePacket}) to derive a reduced one-dimensional Hamiltonian describing the neutrons vertical motion in a Lorentz-violating background as
\begin{align}
H _{z} \equiv \left\langle H _{\mbox{\scriptsize eff}} \right\rangle = \int \psi ^{\ast} (\textbf{r} _{\perp}) \; H _{\mbox{\scriptsize eff}} \; \psi (\textbf{r} _{\perp}) d ^{2} \textbf{r} _{\perp} , \label{RedHamiltonian}
\end{align}
which indeed corresponds to the first order perturbation in the perpendicular $x$-$y$ plane. The rest of this section is devoted to the computation of the reduced Hamiltonian (\ref{RedHamiltonian}).

We first focus on the expectation values of the $\Phi$-independent terms in the Hamiltonian (\ref{EffHam}). From now on, latin indices of the middle of the alphabet $(i,j,k,l)$ refer to the three spatial components $x,y,z$; while the latin indices from the beginning of the alphabet $(a,b,c,e)$ refers to the coordinates $x,y$. We built up to the evaluation of $\left\langle \beta _{i} \hat{p} _{i} \right\rangle$ in two steps. Firstly, we decompose $\beta _{i} \hat{p} _{i}$ into its vertical ($z$) and perpendicular ($x,y$) components by writing $\beta _{i} \hat{p} _{i} = \beta _{a} \hat{p} _{a} + \beta _{z} \hat{p} _{z}$; and secondly we evaluate the expectation value using the Gaussian wave packet (\ref{WavePacket}). The result is
\begin{align}
\left\langle \beta _{i} \hat{p} _{i} \right\rangle = \beta _{a} p _{a} + \beta _{z} \hat{p} _{z} , \label{beta-p}
\end{align}
where we have used that $\left\langle  \hat{p} _{a} \right\rangle = p _{a}$. We can now apply the same procedure to the term $\gamma _{ij} ^{\pm} \hat{p} _{i} \hat{p} _{j}$ to obtain
\begin{align}
\left\langle \gamma _{ij} ^{\pm} \hat{p} _{i} \hat{p} _{j} \right\rangle = \gamma _{ab} ^{\pm} \left\langle \hat{p} _{a} \hat{p} _{b} \right\rangle + ( \gamma _{az} ^{\pm} + \gamma _{za} ^{\pm} ) p _{a} \hat{p} _{z} + \gamma _{zz} ^{\pm} \hat{p} _{z} ^{2} , \label{gamma-pp}
\end{align}
where
\begin{align}
\left\langle \hat{p} _{a} \hat{p} _{b} \right\rangle &= p _{a} p _{b} + \frac{\hbar ^{2}}{2 \sigma ^{2}} \delta _{ab} . \label{pp}
\end{align}
Now we consider the $\Phi$-dependent terms. In the coordinate system attached to the Earth's surface, the Newtonian potential is given by
\begin{align}
\Phi (\textbf{r}) = - \frac{G M _{\oplus}}{r} , \label{NewtonPot}
\end{align}
where $G$ is the gravitational constant, $M _{\oplus}$ is the Earth's mass, and $r ^{2} = x ^{2} + y ^{2} + \left( R _{\oplus} + z \right) ^{2}$, being $R _{\oplus}$ the Earth's radius. Since the potential is not isotropic but axially symmetric, we can use polar coordinates ($x = \rho \cos \varphi$ and $y = \rho \sin \varphi$) to evaluate $\left\langle \Phi \right\rangle$ in the semiclassical state (\ref{WavePacket}), i.e.
\begin{align}
\left\langle \Phi \right\rangle = - \frac{2 G M _{\oplus}}{\sigma ^{2}} \int _{0} ^{\infty} \frac{\rho}{\sqrt{\rho ^{2} + (R _{\oplus} + z ) ^{2}}} e ^{- \frac{\rho ^{2}}{\sigma ^{2}}} d \rho ,
\end{align}
where we have performed the trivial angular integration. The resulting radial integral can be computed in a simple fashion. The final result is
\begin{align}
\left\langle \Phi \right\rangle &= - \frac{\sqrt{\pi} G M _{\oplus}}{\sigma} e ^{\xi ^{2}} \mbox{erfc} (\xi) , \label{Phi}
\end{align}
where $\mbox{erfc} (\xi)$ is the complementary error function \cite{Maths} and $\xi \equiv \left( R _{\oplus} + z \right) / \sigma$. In practice, the experiments with UCNs bouncing on a horizontal mirror are very localized as compared with the Earth's radius, and thus we may approximate the effective potential (\ref{Phi}) for $R _{\oplus} \gg z$ and $R _{\oplus} \gg \sigma$. Using the asymptotic expansion of the complementary error function for large real $x$ \cite{Maths}
\begin{align}
\mbox{erfc} (x) \sim \frac{e ^{-x ^{2}}}{\sqrt{\pi} x} \sum _{n = 0} ^{\infty} (-1) ^{n} \frac{(2n-1)!!}{(2 x ^{2}) ^{n}} , \label{Asymptotic}
\end{align}
we can write the effective potential (\ref{Phi}) as an infinite serie
\begin{align}
\left\langle \Phi \right\rangle = - \frac{G M _{\oplus}}{\sigma \xi} \sum _{n = 0} ^{\infty} (-1) ^{n} \frac{(2n-1)!!}{(2 \xi ^{2}) ^{n}} \equiv \sum _{n = 0} ^{\infty} \left\langle \Phi \right\rangle _{n} .
\end{align}
Since the $n$-th term behaves as $(\sigma / R _{\oplus} ) ^{2n}$, only small values of $n$ contribute. The leading contribution arises from $n = 0$, 
\begin{align}
\left\langle \Phi \right\rangle _{0} = U _{0} + gz , \label{Phi0}
\end{align}
and we can safely disregard the higher order contributions. In this expression, $U _{0} = - G M _{\oplus} / R _{\oplus}$ is the Newtonian potential on the Earth's surface and $g = G M _{\oplus} / R _{\oplus} ^{2}$ is the gravitational acceleration. Equation (\ref{Phi0}) is the expected classical result, and it will be useful to compute the remaining $\Phi$-dependent terms. We can perform an analogous analysis for the term $\beta _{i} \Phi _{, i}$. The axial symmetry of the problem yields to the result $\left\langle \beta _{i} \Phi _{, i} \right\rangle = \beta _{z} \left\langle \Phi \right\rangle _{, z} \equiv \beta _{z} \hat{p} _{z} \left\langle \Phi \right\rangle $. The analysis of the remaining terms, $\gamma _{ij} ^{+} \Phi \hat{p} _{i} \hat{p} _{j}$ and $\gamma _{ij} ^{-} \Phi _{,( i} \hat{p} _{j )}$, is more cumbersome, but it is straightforward. We left the details of the technical computations to the Appendix, and here we only present the final results. The leading order contributions are
\begin{align}
\left\langle \gamma _{ij} ^{\pm} \Phi \hat{p} _{i} \hat{p} _{j} \right\rangle &= \left\langle \Phi \right\rangle _{0} \left\langle \gamma _{ij} ^{\pm} \hat{p} _{i} \hat{p} _{j} \right\rangle , \label{Phi-pp} \\ \left\langle \gamma ^{\pm} _{ij} \Phi _{,(i} \hat{p} _{j)} \right\rangle &= - \delta _{ab} \gamma ^{\pm} _{ab} \frac{g \hbar ^{2}}{4 R _{\oplus}} + \frac{1}{2} \left( \gamma ^{\pm} _{az} + \gamma ^{\pm} _{za} \right) p _{a} \left\langle \Phi \right\rangle _{0,z} \notag \\ & \hspace{2cm} + \gamma ^{\pm} _{zz} \left\langle \Phi \right\rangle _{0,z} \hat{p} _{z} . \label{Phi-dp}
\end{align}
Now we have the pieces to build up the reduced one-dimensional Hamiltonian, which we conveniently write as
\begin{align}
H _{z} = H _{0} + H _{\perp} + V , \label{Red1D-Ham}
\end{align}
where
\begin{align}
H _{0} = \frac{\hat{p} _{z} ^{2}}{2m} + mgz  \label{H0}
\end{align}
is the standard one-dimensional Hamiltonian for a free falling neutron in the absence of Lorentz violation, and
\begin{align}
H _{\perp} = & \, m c ^{2} \left( \alpha ^{-} + \alpha ^{+} \frac{U _{0}}{c ^{2}} \right) + \frac{1}{2m} \left( \gamma _{ab} ^{-} + \gamma _{ab} ^{+} \frac{U _{0}}{c ^{2}} \right) \left\langle \hat{p} _{a} \hat{p} _{b} \right\rangle \notag \\ & - \beta _{a} p _{a} c + \varsigma _{a} p _{a} c \left( 1 + \frac{U _{0}}{c ^{2}} \right) - \frac{1}{2 m c ^{2}} \frac{g \hbar ^{2}}{4 R _{\oplus}} \gamma _{ab} ^{-} \delta _{ab} 
\end{align}
collects constant terms and those depending on the neutron motion in the tangent plane to the Earth's surface. We omit this term as from now since it does not affect the energy eigenvalues measured in the GRANIT experiment. The potential
\begin{align}
V = & \, m g \left[ c _{00} ^{n} + \gamma _{ab} ^{+} \frac{ \left\langle \hat{p} _{a} \hat{p} _{b} \right\rangle}{2 (mc) ^{2}}  \right] z + \left[ \tau ^{-} + \left( 1 + \tau ^{+} \right) \frac{U _{0}}{c ^{2}} \right] \frac{\hat{p} _{z} ^{2}}{2m}  \notag \\ & + \left[ \frac{1 + \tau ^{-}}{2 m c ^{2}} \left\langle \Phi \right\rangle _{0,z} -  \beta _{z} c - \left( 1 + \frac{U _{0}}{c ^{2}} \right) \frac{\zeta _{a} p _{a}}{m} \right] \hat{p} _{z} \notag \\ & + \frac{\varsigma _{a} p _{a}}{c} g z + \varsigma _{z} \left( 1 + \frac{U _{0}}{c ^{2}} \right) \hat{p} _{z} c + \frac{\varsigma _{z}}{c} \left( g z \hat{p} _{z} + \frac{1}{2} \left\langle \Phi \right\rangle _{,z} \right) \notag \\ & - \frac{\zeta _{a} p _{a}}{2 m c ^{2}} \left( \left\langle \Phi \right\rangle _{0,z} + 2 g z \hat{p} _{z} \right)  + \frac{g \left( 1 + \tau ^{+} \right)}{2m c ^{2}} z \hat{p} _{z} ^{2}  \label{V-Pot}
\end{align}
is the one which has possibilities of affecting the neutrons vertical motion. In this expression we have defined
\begin{align}
\zeta _{a} \equiv - ( \gamma _{az} ^{\pm} + \gamma _{za} ^{\pm} ) / 2 \quad , \quad \tau ^{\pm} \equiv \gamma _{zz} ^{\pm} - 1 .
\end{align}
As we shall see in the next section, many of these terms do not contribute to the energy shifts.

\section{Energy shifts and bounds on $c _{\mu \nu} ^{n}$ SME coefficients}
\label{BoundSec}

In this section we will work out the energy shifts on the neutron states due to the SME terms and we will compare our theoretical results with the experimental ones obtained in the GRANIT experiment. This comparison will allow us to establish a simple formula for the upper bound on the SME coefficients as a function of the maximal experimental uncertainty. We first describe in short the  the neutron states in the absence of the SME.

The wave function of a quantum bouncer obeys the stationary Schr\"{o}dinger equation for the vertical motion along the $z$ axis: $H _{0} \psi = E \psi$, with the Hamiltonian given by Eq. (\ref{H0}). The solution must obey the following boundary conditions: $\psi (z)$ must vanish asymptotically as $z \rightarrow \infty$, and $\psi (z = 0) = 0$ because of the presence of a mirror at $z = 0$. The general solution of the eigenvalue equation can be written in terms of the Airy functions $\mbox{Ai}$ and $\mbox{Bi}$ \cite{Vallee}. Since the latter goes to infinity as its argument grows, it is not an acceptable solution for this problem. The appropriate normalized solution is found to be
\begin{align}
\psi _{n} (z) = \frac{1}{\sqrt{l _{0}}} \frac{\mbox{Ai} (a _{n} + z/l _{0} )}{\mbox{Ai} ^{\prime} (a _{n})} \Theta (z) , \label{UnpWaveFunc}
\end{align}
where $a _{n}$ is the $n$-th zero of the Airy function $\mbox{Ai}$, $l _{0} = \sqrt[3]{\hbar ^{2} / (2m ^{2}g ) }$ is the gravitational length and $\Theta (z)$ is the Heaviside function. The boundary condition at $z=0$ defines the quantum state energies
\begin{align}
E _{n} = - mgl_{0} a _{n} . \label{UnpEnergy}
\end{align}
Within the classical description, a neutron with energy $E _{n}$ can rise in the gravitational field up to the height $h _{n} = E _{n} / mg = - a _{n} l _{0}$. The heights for the two lowest quantum states are \cite{GRANIT}
\begin{equation}
h _{1} = 13.7 \mu \mbox{m} \qquad , \qquad h _{2} = 24.0 \mu \mbox{m} . \label{HeightsTheo}
\end{equation}
Because of the weakness of the gravitational interaction and the number of systematic errors in laboratory conditions, quantum states in a gravitational field have been hardly detected. In spite of these difficulties, the GRANIT experiment has recently confirmed the quantum-mechanical prediction that a non coherent beam of UCNs propagating upwards in the Earth's gravity field reach quantized heights only. The experimental average values of the two lowest critical heights (taken from \cite{Nesvizhevsky3}) are
\begin{align}
h _{1} ^{\mbox{\scriptsize exp}} &= \left( 12.2 \pm 1.8 _{\scriptsize \mbox{sys}} \pm 0.7 _{\scriptsize \mbox{stat}} \right) \mu \mbox{m} , \notag \\ h _{2} ^{\mbox{\scriptsize exp}} &= \left( 21.6 \pm 2.2 _{\scriptsize \mbox{sys}} \pm 0.7 _{\scriptsize \mbox{stat}} \right) \mu \mbox{m} . \label{HeightsExp}
\end{align}
The theoretical values are therefore located within the error bars. As a consequence of the good agreement between theory and experiment, this finding could be used for bounding deviations from the standard theory due to an eventual new physical mechanism. It has been used, for example, to constraint short-range gravitational interactions \cite{ShortRange}, axion-like interactions \cite{Axion} and the fundamental length scale in polymer quantum mechanics \cite{Martin}. In the problem at hand, the potential $V$ given by Eq. (\ref{V-Pot}), will cause small shifts $\Delta E _{n}$ in the neutron energy spectrum which must satisfy the constraint
\begin{equation}
\vert \Delta E _{n} \vert < \vert \Delta E _{n} ^{\scriptsize \mbox{exp}} \vert , \label{BoundEnergy}
\end{equation}
where $\vert \Delta E _{n} ^{\scriptsize \mbox{exp}} \vert$ is the maximal experimental error. Explicitly, the energy shifts can be worked out using the formalism of nondegenerate perturbation theory on the wave functions $\psi _{n} (z)$ up to linear order in the SME coefficients, that is: $\Delta E _{n} = \left\langle V \right\rangle = \int \psi _{n} ^{\ast} V \psi _{n} dz$. Using the properties of the Airy functions \cite{Vallee}, one can derive the following results
\begin{align}
\left\langle p _{z} \right\rangle = 0 , \quad mg \left\langle z \right\rangle = \frac{2}{3} E _{n} , \quad \left\langle \hat{p} _{z} ^{2} / 2m \right\rangle = \frac{1}{3} E _{n} , \notag \\ \quad \frac{g}{2mc ^{2}} \left\langle z \hat{p} _{z} ^{2} \right\rangle = - \frac{2}{15} a _{n} E _{n} \frac{g l _{0}}{c ^{2}} ,
\end{align}
which yields the energy shifts
\begin{align}
\frac{\Delta E _{n}}{E _{n}} = & \frac{1}{3} \left( 2 c _{00} ^{n} + \tau ^{-} \right) + \frac{1}{3} \left( 1 + \tau ^{+} \right) \left( \frac{U _{0}}{c ^{2}} + \frac{2}{5} \frac{E _{n}}{m c ^{2}} \right) \notag \\ &  + \frac{1}{3} \gamma _{ab} ^{+} \left( \frac{v _{a} v _{b}}{c ^{2}} + \frac{\hbar ^{2}}{2 m ^{2} c ^{2} \sigma ^{2}} \delta _{ab} \right) + \frac{2}{3} \varsigma _{a} \frac{v _{a}}{c} , \label{EnergyShifts}
\end{align}
where we have used that $p _{a} = m v _{a}$, being $v _{a}$ the neutrons velocity. For nonrelativistic neutrons in low quantum states, we find that $U _{0} / c ^{2} \approx 10 ^{-10}$, $E _{n} / (m c ^{2}) \approx 10 ^{-22}$, $v _{a} / c \approx 10 ^{-7}$ and $\hbar ^{2} / ( m ^{2} c ^{2} \sigma ^{2} ) \approx 10 ^{-15}$, and thus we can disregard the terms involving products of SME coefficients and these quantities.  Therefore we are left with 
\begin{align}
\frac{\Delta E _{n}}{E _{n}} = & \frac{1}{3} \left( 2 c _{00} ^{n} + \tau ^{-} \right) , \label{EnergyShifts2}
\end{align}
which after substitution into Eq. (\ref{BoundEnergy}) produces
\begin{equation}
\vert c _{00} ^{n} - 2 c _{zz} ^{n} \vert < 3 \frac{\vert \Delta E _{n} ^{\scriptsize \mbox{exp}} \vert}{E _{n}} . \label{BoundSME}
\end{equation}
For the first two lowest quantum states, we know that $\vert \Delta E _{1} ^{\scriptsize \mbox{exp}} \vert = 0.102  \mbox{peV}$ and $\vert \Delta E _{2} ^{\scriptsize \mbox{exp}} \vert = 0.051 \mbox{peV}$ \cite{Nesvizhevsky3}. With these values Eq. (\ref{BoundSME}) yields the constraint $\vert c _{00} ^{n} - 2 c _{zz} ^{n} \vert < 10 ^{-2}$. Of course, this bound is largely far from the expected values for the SME coefficients, but it can compete with current bounds with an improvement of the experimental precision in the measurement of the quantum states of UCNs in a gravitational field, as we will discuss in the next section. According to the current data tables for the SME coefficients \cite{table1}, we observe that there are only very few bounds involving the $c _{00} ^{n}$ and $c _{zz} ^{n}$ coefficients. Even more, the combination $\vert c _{00} ^{n} - 2 c _{zz} ^{n} \vert$ which emerges in this work has not been reported.

We point out that although the coefficients $a _{i} ^{n}$ and $e _{i} ^{n}$ appear explicitly in the energy shifts (\ref{EnergyShifts}), they are suppressed by the additional factor $v _{a}/c$, and thus the experimental precision leads to the noncompetitive bound $ a _{i} ^{n} /m - e _{i} ^{n} < 10 ^{5}$. The suppression of the observable effects of the combination $\varsigma _{i} =  a _{i} ^{n} /m - e _{i} ^{n}$ by the factor $v _{i}/c$ deserves some explanation. If we look at the Lagrangian density in Eq. (\ref{Action}), the contributions proportional to $a _{i} ^{n}$ and $e _{i} ^{n}$ are not suppressed. Nonrelativistically, however, they correspond to a different order of approximation in the FW transformation than that of the coefficients $c _{jk} ^{n}$ and $c _{00} ^{n}$, as can be seen in the Hamiltonian (\ref{Schrodinger}). Indeed, the coefficients $c _{jk} ^{n}$ and $c _{00} ^{n}$ are of the order $p^{2}/m$, while the combination $\varsigma _{i} =  a _{i} ^{n} /m - e _{i} ^{n}$ is of the order $pc = (p ^{2}/m) (v/c)^{-1}$, thus revealing the nature of the additional factor of the latter. It is worth to mention that, even though we got a large value for the bound of the coefficients $a _{i} ^{n}$ and $e _{i} ^{n}$, thus far such coefficients are unbounded by experiment in the neutron sector. In this manner, the present work, beyond of theoretical importance, can provide novel bounds in the context of the SME. To obtain a significant result it is necessary to reduce the value of such a bound by some orders of magnitude, which indeed is possible as we will discuss later. 

Bounds on the $c _{\mu \nu} ^{n}$ neutron sector coefficients of the SME have been reported by using different physical systems. For example, gravimetry sets the bounds $c _{TJ} ^{n} < 10 ^{-5}$, with $J=X,Y,Z$ \cite{Gravimetry}. Similarly, nuclear binding models and Cs interferometers yield $c _{TT} ^{n} < 10 ^{-6}$ \cite{interferometers}. More stringent bounds on specific combinations of the neutron $c _{\mu \nu} ^{n}$ coefficients come from pulsar timing, namely, $\mbox{min} \left( \vert c _{11} ^{n} - c _{22} ^{n} \vert , \vert c _{11} ^{n} - c _{33} ^{n} \vert , \vert c _{22} ^{n} - c _{33} ^{n} \vert \right) < 1.7 \times 10 ^{-8}$ \cite{Altschul}. Importantly, the best current bounds on such coefficients come from the $^{21}$Ne-Rb-K comagnetometer, which constrain the combinations $c _{YZ} ^{n} + c _{ZY} ^{n}$, $c _{XZ} ^{n} + c _{ZX} ^{n}$, $c _{XY} ^{n} + c _{YX} ^{n}$ and $c _{XX} ^{n} - c _{YY} ^{n}$ at a level of $10 ^{-29}$ \cite{Comagnetometer}. It is worth to mention that, however, none of these experiments provide bounds on the combination $\vert c _{00} ^{n} - 2 c _{zz} ^{n} \vert$, which is the one obtained here. In the next section we will discuss two sensitive experiments which also involve quantum states of UCNs in the Earth's gravity field and which would improve by some orders of magnitude our current bound.

\section{Discussion and outlook}
\label{Summary}

The experimental physics of slow neutrons has undergone significant evolution in the last decades. Recent high-sensitivity experiments, called GRANIT, performed by V. V. Nesvizhevsky \textit{et al.} at the Institute Laue-Langevin, show that UCNs in the Earth's gravitational field move at certain well-defined (quantized) heights, in agreement with quantum mechanical predictions. Due to the good agreement between theory and experiment, neutron gravitational quantum states can be used for constraining deviations from the standard theory due to eventual new physical mechanisms. In light of this, in this paper we have investigated how the fermion sector of the SME affects the gravitational quantum states of UCNs, mainly focusing on the energy shifts. 

We first consider the Dirac equation in a Newtonian field, which is appropriate to characterize the gravitational field in any terrestrial experiment. Since UCN systems are nonrelativistic, we have used the nonrelativistic limit of the Dirac equation which can be obtained by using the Foldy-Wouthuysen procedure. GRANIT experiments use an intense horizontal beam of unpolarized UCNs directed slightly upwards and allowing the neutrons to fall onto a horizontal mirror, and then the neutrons horizontal motion is governed by classical laws, while the vertical motion is quantized. In order to isolate the effects along the axis of free fall, we have considered a semiclassical Gaussian wave packet and then obtained a reduced Hamiltonian by computing the expectation value on the perpendicular axes. We find a $z$- and $\hat{p} _{z}$-dependent perturbative potential $V$, given by Eq. (\ref{V-Pot}), which is proportional to the SME coefficients and the (zeroth order) Newtonian gravity field $\left\langle \Phi \right\rangle _{0} = U _{0} + gz$. We have worked out the energy shifts $ \Delta E = \left\langle V \right\rangle $ to first order in perturbation theory, and we found it contains both SME- and relativistic-corrections. Next we used the maximal experimental precision of the GRANIT experiment to set bounds to the SME coefficients. The lowest quantum states of UCNs yields $\vert c _{00} ^{n} - 2 c _{zz} ^{n} \vert < 10 ^{-2}$, which although is far from the current bounds obtained using other physical systems (e.g. by gravimetry, nuclear binding models, Cs interferometry, pulsar timing and $^{21}$Ne-Rb-K comagnetometer), it opens a new window to test Lorentz- and CPT-violation using UCN systems. It is worth mentioning that the specific combination we find, $c _{00} ^{n} - 2 c _{zz} ^{n}$, has not been constrained by any of the aforementioned experiments, thus justifying the importance of the present work in the Standard-Model Extension framework. Even more, as we can see in the current data tables \cite{table1}, there exist no bounds on the $a _{i} ^{n}$ and $e _{i} ^{n}$ coefficients in the minimal neutron sector of the SME, so far. In the present work, indeed, we find that the GRANIT experiment is sensitive to these coefficients, however it does not provide a realistic bound for them. This is so because in the nonrelativistic Hamiltonian (\ref{Schrodinger}), the coefficients $a _{i} ^{n}$ and $e _{i} ^{n}$ belong to a different order of approximation than that of the $c _{00} ^{n}$ and $c _{jk} ^{n}$ coefficients in the Foldy-Wouthuysen procedure, thus justifying the additional suppression of their bound by the factor $v/c$. An interesting improvement of our bounds can be achieved with other experiments involving UCNs: gravity-resonance-spectroscopy and neutron whispering gallery wave. Although they are beyond the scope of this paper, we will briefly discuss how these experiments can enhance the bounds on the SME coefficients.

An interesting feature of the quantum bouncer, in contrast to the harmonic oscillator, is the fact that levels are not equidistant in energy. Therefore a combination of any two states can be treated as a two level system. This fact has been used by T. Jenke and colleagues to induce transitions between the $n =1$ and $n=3$ states by means of mechanical oscillations of the mirror \cite{Jenke}. This new spectroscopic technique is called gravity-resonance-spectroscopy. In the experiment, the statistical sensitivity of the energy difference between states $\left| 1 \right\rangle$ and $\left| 3 \right\rangle$ is $7.6 \times 10 ^{-3}$, which corresponds to an uncertainty in energy of $\delta E = 2 \times 10 ^{-14}$eV. Ignoring the nonzero transitions induced by the Lorentz violating perturbation (\ref{V-Pot}), a rough estimation yields an improvement of one order of magnitude on the bound (\ref{BoundSME}) for the $c _{\mu \nu} ^{n}$ coefficients. This, of course, requires a detailed theoretical analysis which we leave for a future investigation.

In Sec. \ref{BoundSec} we have derived an expression for the upper bound on the $c _{\mu \nu} ^{n}$ SME coefficients in terms of the experimental precision $\Delta E _{n} ^{\scriptsize \mbox{exp}}$ and the unperturbed energy levels $E _{n}$, from which we learn that a better bound can be obtained by improving the experimental precision and/or by considering a system in which the unperturbed energy levels be considerably greater than those of the quantum bouncer. This leads us to consider the recently observed neutron centrifugal states \cite{Centrifugal}, which is the quantum analog of the so-called whispering gallery wave. In this case, UCNs are scattered by a perfect cylindrical mirror with a radius of a few centimeters, in which neutrons are affected by a huge centrifugal accelerations of the order $10 ^{5}$-$10 ^{7} g$. Most neutrons entering at a tangential trajectory are deviated to small angles. However, some neutrons are captured into long-living centrifugal states which behaves exactly as the neutron gravitational quantum states discussed in this paper. The fundamental difference is that in the former case the centrifugal force plays the role of gravity, while in the latter we refer to the well-worked Newtonian gravity field. A rough calculation shows that the characteristic energy scale is of the order of neV, which together with the considered experimental precision $10 ^{-2}$peV, could improves our upper bound by $5$ orders of magnitude. This is an interesting system which deserves a rigorous investigation.

\acknowledgments
C. A. E. and A. M. R were supported by CONACyT postdoctoral grants No. 234745 and  No. 234774, respectively. We greatly appreciate correspondence with Professor V. V. Nesvizhevsky. Helpful discussions with Professor R. Potting are warmly appreciated. We thank the referee for his/her comments and suggestions which have substantially improved the scope of this work.

\appendix

\section{Expectation values}

In this section we evaluate the expectation value of the terms $\gamma _{ij} ^{\pm} \Phi \hat{p} _{i} \hat{p} _{j}$ and $\gamma _{ij} ^{\pm} \Phi _{,( i} \hat{p} _{j )}$, as defined in Eq. (\ref{RedHamiltonian}). Decomposing the former into its vertical and perpendicular components, its expectation value can be written as
\begin{align}
\left\langle \gamma _{ij} ^{\pm} \Phi \hat{p} _{i} \hat{p} _{j} \right\rangle &= \gamma _{ab} ^{\pm} \left\langle \Phi \hat{p} _{a} \hat{p} _{b} \right\rangle + ( \gamma _{az} ^{\pm} + \gamma _{za} ^{\pm} ) \left\langle \Phi \hat{p} _{a} \right\rangle \hat{p} _{z} \notag \\ & \hspace{3cm} + \gamma _{zz} ^{\pm} \left\langle \Phi \right\rangle \hat{p} _{z} ^{2} . \label{Phi-pp-Ap}
\end{align}
Now we evaluate each term involved. We start with
\begin{align}
\left\langle \Phi \hat{p} _{a} \right\rangle &= - i \hbar \int \Phi (\textbf{r}) \left( i \frac{p _{a}}{\hbar} - \frac{x _{a}}{\sigma ^{2}} \right) \vert \psi (\textbf{r} _{\perp}) \vert ^{2} d ^{2} \textbf{r} _{\perp} . \label{Phi-p-Ap}
\end{align}
Using the axial symmetry of the gravitational potential and that of the wave packet, we easily find
\begin{align}
\left\langle \Phi \hat{p} _{a} \right\rangle &= p _{a} \int \Phi (\textbf{r}) \vert \psi (\textbf{r} _{\perp}) \vert ^{2} d ^{2} \textbf{r} _{\perp} = p _{a} \left\langle \Phi \right\rangle , \label{Phi-pa-Sec}
\end{align}
where $\left\langle \Phi \right\rangle$ was computed in the main text. The next term to be considered is
\begin{align}
\left\langle \Phi \hat{p} _{a} \hat{p} _{b} \right\rangle &= - \hbar ^{2} \int \Phi (\textbf{r}) \psi ^{\ast} (\textbf{r} _{\perp}) \frac{\partial ^{2}}{\partial x ^{a} \partial x ^{b}} \psi (\textbf{r} _{\perp}) d ^{2} \textbf{r} _{\perp} .
\end{align}
Taking the derivatives of the wave packet (\ref{WavePacket}) and using the axial symmetry of the problem this expression can be written as
\begin{align}
\left\langle \Phi \hat{p} _{a} \hat{p} _{b} \right\rangle = & \left( p _{a} p _{b} + \frac{\hbar ^{2}}{\sigma ^{2}} \delta _{ab} \right) \left\langle \Phi \right\rangle \notag \\ & - \frac{\hbar ^{2}}{\sigma ^{4}} \int \Phi (\textbf{r}) x _{a} x _{b} \vert \psi (\textbf{r} _{\perp}) \vert ^{2} d ^{2} \textbf{r} _{\perp} , \label{Phi-pp-App2}
\end{align}
where $\left\langle \Phi \right\rangle$ is given by Eq. (\ref{Phi}). The second term in Eq. (\ref{Phi-pp-App2}), to be called $Q _{ab}$ for brevity, must be computed explicitly. We first observe that the integral is nonzero only for $a = b$. Using polar coordinates ($x = \rho \cos \phi$ and $y = \rho \sin \phi$) and performing the trivial angular integration, the function $Q _{ab}$ becomes
\begin{align}
Q _{ab} &= \frac{GM _{\oplus} \hbar ^{2}}{\sigma ^{6}} \delta _{ab} \int _{0} ^{\infty} \frac{\rho ^{3}}{r} e ^{- \frac{\rho ^{2}}{\sigma ^{2}}} d \rho .
\end{align}
With the simple change of variables $\lambda = r / \sigma$, this integral can be brought to the simple form
\begin{align}
Q _{ab} &= \frac{G M _{\oplus} \hbar ^{2}}{\sigma ^{3}} \delta _{ab} \, e ^{\xi ^{2}} \int _{\xi} ^{\infty} \left( \lambda ^{2} - \xi ^{2} \right) e ^{- \lambda ^{2}} d \lambda ,
\end{align}
which can be easily evaluated to obtain
\begin{align}
Q _{ab} = \frac{G M _{\oplus} \hbar ^{2}}{4 \sigma ^{3}} \delta _{ab} \left[ 2 \xi + \sqrt{\pi } \left(1-2 \xi ^{2} \right) e ^{\xi ^{2}} \mbox{erfc}(\xi) \right] , \label{inter}
\end{align}
where $\xi = \left( R _{\oplus} + z \right) / \sigma$. Since $R _{\oplus} \gg z \gg \sigma$, we have to consider the asymptotic behavior of Eq. (\ref{inter}) for $\xi \gg 1$. Using Eq. (\ref{Asymptotic}) up to second order we finally obtain
\begin{align}
Q _{ab} & \sim \frac{G M _{\oplus} \hbar ^{2}}{2 \xi \sigma ^{3}} \delta _{ab} \approx - \frac{\hbar ^{2}}{2 \sigma ^{2}} \left\langle \Phi \right\rangle _{0} \delta _{ab} , \label{Qz}
\end{align}
where $\left\langle \Phi \right\rangle _{0}$ is the leading order of the Newtonian potential given by Eq. (\ref{Phi0}). The substitution of this result into Eq. (\ref{Phi-pp-App2}) then produces
\begin{align}
\left\langle \Phi \hat{p} _{a} \hat{p} _{b} \right\rangle &= \left( p _{a} p _{b} + \frac{\hbar ^{2}}{\sigma ^{2}} \delta _{ab} \right) \left\langle \Phi \right\rangle - \frac{\hbar ^{2}}{2 \sigma ^{2}} \left\langle \Phi \right\rangle _{0} \delta _{ab} , \label{Phi-pp-App3}
\end{align}
and the leading order of this result establishes Eq. (\ref{Phi-pp}).

Now we evaluate $\left\langle \gamma _{ij} ^{\pm} \Phi _{,( i} \hat{p} _{j )} \right\rangle$. We proceed first by decomposing one of this terms into its vertical and perpendicular components:
\begin{align}
\left\langle \gamma ^{\pm} _{ij} \Phi _{,i} \hat{p} _{j} \right\rangle &= \gamma ^{\pm} _{ab} \left\langle \Phi _{,a} \hat{p} _{b} \right\rangle + \gamma ^{\pm} _{az} \left\langle \Phi _{,a} \right\rangle \hat{p} _{z} + \gamma ^{\pm} _{za} \left\langle \Phi _{,z} \hat{p} _{a} \right\rangle \notag \\ & \hspace{3cm} + \gamma ^{\pm} _{zz} \left\langle \Phi \right\rangle _{,z} \hat{p} _{z} . \label{Phi-p-symm}
\end{align}
Some simplifications occur in this expression. First, we observe that $\left\langle \Phi _{,z} \right\rangle = \left\langle \Phi \right\rangle  _{,z}$, which follows from the fact that we can commute a $z$-derivative with an integral over the perpendicular coordinates $x$ and $y$. In a similar fashion we obtain $\left\langle \Phi _{,z} \hat{p} _{a} \right\rangle = p _{a} \left\langle \Phi \right\rangle _{,z}$, where we have used both the fact that $\partial _{z}$ and $\int d ^{2} \textbf{r} _{\perp}$ commute, and the result of Eq. (\ref{Phi-pa-Sec}). Also, since the probability density $\vert \psi (\textbf{r} _{\perp}) \vert ^{2}$ is axially symmetric and $\Phi _{,a} \propto x _{a} / r ^{3}$, then the expectation value $\left\langle \Phi _{,a} \right\rangle$ is identically zero. Thus we are left with $\left\langle \Phi _{,a} \hat{p} _{b} \right\rangle$. Taking the required derivatives, this term can be explicitly written as
\begin{align}
\left\langle \Phi _{,a} \hat{p} _{b} \right\rangle &= - G M _{\oplus} \hbar ^{2} \int \frac{x _{a}}{r ^{3}} \left( i \frac{p _{b}}{\hbar} - \frac{x _{b}}{\sigma ^{2}} \right) \vert \psi (\textbf{r} _{\perp}) \vert ^{2} d ^{2} \textbf{r} _{\perp} .
\end{align}
The first integral vanishes by symmetry considerations, while the second one is nonzero only for $a = b$. Using polar coordinates and performing the angular integration we get
\begin{align}
\left\langle \Phi _{,a} \hat{p} _{b} \right\rangle &= \frac{G M _{\oplus} \hbar ^{2}}{\sigma ^{4}} \delta _{ab} \int _{0} ^{\infty} \frac{\rho ^{3}}{r ^{3}} e ^{- \frac{\rho ^{2}}{\sigma ^{2}}} d \rho ,
\end{align}
which can be cast into a more simple form introducing the change of variables $\lambda = r / \sigma$:
\begin{align}
\left\langle \Phi _{,a} \hat{p} _{b} \right\rangle &=  \frac{G M _{\oplus} \hbar ^{2}}{\sigma ^{3}} \delta _{ab} \, e ^{\xi ^{2}} \int _{\xi} ^{\infty} \frac{\lambda ^{2} - \xi ^{2}}{\lambda ^{2}} e ^{- \lambda ^{2}} d \lambda .
\end{align}
As in the previous cases, the resulting integral can be expressed in terms of the complementary error function:
\begin{align}
\left\langle \Phi _{,a} \hat{p} _{b} \right\rangle &= \frac{GM _{\oplus} \hbar ^{2}}{\sigma ^{3}} \delta _{ab}\left[ - \xi + \frac{\sqrt{\pi}}{2} \left( 1 + 2 \xi ^{2} \right) e ^{\xi ^{2}}  \mbox{erfc} ( \xi ) \right] ,
\end{align}
from which, with the help of Eq. (\ref{Asymptotic}), we extract its asymptotic behavior for $\xi \gg 1$ to finally obtain
\begin{align}
\left\langle \Phi _{,a} \hat{p} _{b} \right\rangle & \sim - \frac{GM _{\oplus} \hbar ^{2}}{4 \sigma ^{3} \xi ^{3}} \delta _{ab} \approx - \frac{g \hbar ^{2}}{4 R _{\oplus}} \delta _{ab} \left( 1 - 3 \frac{z}{R _{\oplus}} \right) ,
\end{align}
where in the last approximation we have used that $z \ll R _{\oplus}$. Note that the second term is strongly suppressed with respect to the first one, and thus we can ignore it. Inserting this result into Eq. (\ref{Phi-p-symm}) and symmetrizing it we obtain,
\begin{align}
\left\langle \gamma ^{\pm} _{ij} \Phi _{,(i} \hat{p} _{j)} \right\rangle &= - \delta _{ab} \gamma ^{\pm} _{ab} \frac{g \hbar ^{2}}{2 R _{\oplus}} + \frac{1}{2} \left( \gamma ^{\pm} _{az} + \gamma ^{\pm} _{za} \right) p _{a} \left\langle \Phi \right\rangle _{,z} \notag \\ & \hspace{3cm} + \gamma ^{\pm} _{zz} \left\langle \Phi \right\rangle _{,z} \hat{p} _{z} .
\end{align}
The leading order of this result establishes Eq. (\ref{Phi-dp}).

\end{document}